\documentclass[twocolumn,showpacs,amsmath,amssymb,prb,superscriptaddress]{revtex4}

\usepackage{graphicx}
\usepackage{dcolumn}
\usepackage{bm}
\usepackage{color}
\usepackage{ulem}
\usepackage{amsfonts,amsthm}

\begin{document}

\newcommand{\bR}{\mbox{\boldmath $R$}}
\newcommand{\tr}[1]{\textcolor{red}{#1}}
\newcommand{\trs}[1]{\textcolor{red}{\sout{#1}}}
\newcommand{\tb}[1]{\textcolor{blue}{#1}}
\newcommand{\tbs}[1]{\textcolor{blue}{\sout{#1}}}
\newcommand{\Ha}{\mathcal{H}}
\newcommand{\mh}{\mathsf{h}}
\newcommand{\mA}{\mathsf{A}}
\newcommand{\mB}{\mathsf{B}}
\newcommand{\mC}{\mathsf{C}}
\newcommand{\mS}{\mathsf{S}}
\newcommand{\mU}{\mathsf{U}}
\newcommand{\mX}{\mathsf{X}}
\newcommand{\sP}{\mathcal{P}}
\newcommand{\sL}{\mathcal{L}}
\newcommand{\sO}{\mathcal{O}}
\newcommand{\la}{\langle}
\newcommand{\ra}{\rangle}
\newcommand{\ga}{\alpha}
\newcommand{\gb}{\beta}
\newcommand{\gc}{\gamma}
\newcommand{\gs}{\sigma}
\newcommand{\vk}{{\bm{k}}}
\newcommand{\vq}{{\bm{q}}}
\newcommand{\vR}{{\bm{R}}}
\newcommand{\vQ}{{\bm{Q}}}
\newcommand{\vga}{{\bm{\alpha}}}
\newcommand{\vgc}{{\bm{\gamma}}}
\newcommand{\Ns}{N_{\text{s}}}


\title{
First-principles calculation of transition-metal impurities in LaFeAsO
}

\author{Kazuma Nakamura}
\email{kazuma@solis.t.u-tokyo.ac.jp}
\affiliation{%
Department of Applied Physics, University of Tokyo, and JST TRIP/CREST,
7-3-1 Hongo, Bunkyo-ku, Tokyo, 113-8656, Japan}%

\author{Ryotaro Arita}%
\affiliation{%
Department of Applied Physics, University of Tokyo, and JST TRIP/CREST, 
7-3-1 Hongo, Bunkyo-ku, Tokyo, 113-8656, Japan}%

\author{Hiroaki Ikeda}%
\affiliation{%
Department of Physics, Kyoto University, and JST TRIP,  
Sakyo-ku, Kyoto, 606-8502, Japan}%

\date{\today}

\begin{abstract}
We present a systematic {\it ab initio} study based on density-functional calculations to understand impurity effects in iron-based superconductors. Effective tight-binding Hamiltonians for the $d$-bands of LaFeAsO with various transition-metal impurities such as Mn, Co, Ni, Zn, and Ru are constructed using maximally-localized Wannier orbitals. Local electronic structures around the impurity are quantitatively characterized by their onsite potential and transfer hoppings to neighboring sites. We found that the impurities are classified into three groups according to the derived parameters: For Mn, Co, and Ni, their impurity-3$d$ levels measured from the Fe-3$d$ level are nearly $0.3$ eV, $-0.3$ eV, and $-0.8$ eV, respectively, while, for the Zn case, the $d$ level is considerably deep as $- 8$ eV. For the Ru case, although the onsite-level difference is much smaller as {\it O}(0.1) eV, the transfer integrals around the impurity site are larger than those of the pure system by 20\% $\sim$ 30\%, due to the large spatial spread of the Ru-4$d$ orbitals. We also show that, while excess carriers are tightly trapped around the impurity site (due to the Friedel sum rule), there is a rigid shift of band structure near the Fermi level, which has the same effect as carrier doping.
\end{abstract}

\maketitle
\section{Introduction}
Since the discovery of high transition-temperature superconductivity in F-doped LaFeAsO by Kamihara {\it et al.},~\cite{Kamihara} iron-based superconductors have attracted an enormous interest. While extensive theoretical and experimental studies have been performed so far, the pairing mechanism is yet to be unclear. Among various physical properties characterizing the new superconductors, the robustness of the superconductivity against impurity doping has been expected to provide a crucial hint to determine the symmetry of the pairing gap function and pin down precisely the glue of the pairing.

Up to present, several types of the gap symmetry has been proposed.~\cite{rf:Mazin,rf:Kuroki,rf:Ikeda,rf:Ikead2,rf:Graser,rf:Nomura,rf:FWang,rf:Stanscu,rf:JZhang,rf:Yanagi,rf:Yanagi2,rf:Kontani} 
In particular, the $s_{++}$-wave pairing without sign changes in the Brillouin zone and the $s_{\pm}$-wave pairing for which electron and hole Fermi pockets have opposite sign have been recognized as strong candidates.
Experimentally, although some phase-sensitive experiments suggest the existence of sign changes in the gap function,~\cite{STM,halfflux} there have been many reports which claim that the pair-breaking effect by impurity doping is rather weak for various impurity dopants. Especially, Ni {\it et al.}~\cite{Canfield} and Kobayashi {\it et al.}~\cite{Sato} found that the reduction of the transition temperature is simply determined by the amount of doped carriers, but not by the species of impurity dopants or the details of the local electronic structure around the dopant. 
While these observations can be understood in terms of the $s_{++}$-pairing scenario, it is not trivial whether the $s_{\pm}$-pairing can also explain the experiments. Indeed, according to theoretical calculations for tight-binding impurity models with the T-matrix approximation~\cite{OnariKontani} or the Bogoliubov-de Genne equation,~\cite{KariyadoOgata} very small amount of impurities can easily wash out the $s_{\pm}$ pairing, especially when the onsite impurity potential is positive. On the other hand, within the same models, we may still think about the possibility of the $s_{\pm}$ pairing when the onsite impurity potential is negative and not so deep.~\cite{OnariKontani,KariyadoOgata} Thus, quantitative and realistic estimations of the impurity potential are highly desired.

In fact, there have been several {\it ab initio} calculations studying the impurity effects in iron-based superconductors.~\cite{Kemper,Zhang_1,Zhang_2,Wadati} Among them, Kemper {\it et al.} estimated the impurity potential of Co in BaFe$_2$As$_2$.~\cite{Kemper} However, one of the important aspects of the impurity effect in the iron-based superconductors is their appreciable impurity-species dependence,~\cite{Canfield,Sato,Muromachi} so that we need a systematic calculation for various dopants. 

The purpose of the present paper is to study systematically local electronic structures of various non-magnetic impurities such as Mn, Co, Ni, Zn, and Ru, by means of {\it ab initio} density-functional calculations.~\cite{Ref_DFT} For quantitative characterization of the impurity effect, we derive  tight-binding impurity models using maximally-localized Wannier orbitals.~\cite{Marzari,Souza} We see that the local electronic structure around the impurity can be classified into three types; (i) for Mn, Co, and Ni, their $d$ levels measured from the Fe-3$d$ level are $0.3$ eV, $-0.3$ eV, and $-0.8$ eV, respectively, (ii) the Zn-3$d$ level lies very deep ($\sim -8$ eV), so that the Zn site can be regarded as a simple vacancy, and (iii) while the level difference between Ru-4$d$ and Fe-3$d$ orbitals is quite small as {\it O}(0.1) eV, there are appreciable transfer modulations. We also show that the impurity substitution indeed work as an effective doping, even though excess carriers are tightly trapped at the impurity site. 

The structure of the paper is as follows: In Sec. II, we derive tight-binding impurity models with maximally-localized Wannier functions and introduce onsite-potential differences and transfer modulations to characterize the local electronic structure around the impurity. Section~III is devoted to present the computational results for supercell-band structures, basic properties of the Wannier functions, and parameters in the tight-binding model. In Sec.~IV, we discuss how impurity doping affects low-energy electronic structures of LaFeAsO. The concluding remarks are given in Sec.~V.     

\section{Basic quantities specifying an impurity Hamiltonian}
We consider the following effective one-body Hamiltonian including 
a single impurity as 
\begin{eqnarray}
\mathcal{H}
&=& \sum_{\sigma} \sum_{i\ne0} \sum_{\mu} 
         \epsilon_{\mu i}  
         a_{\mu i}^{\sigma \dagger} 
         a_{\mu i}^{\sigma} \nonumber \\
&+& \sum_{\sigma} \sum_{i < j \ne 0} \sum_{\mu\nu}  
  \bigr( t_{\mu i \nu j}  
         a_{\mu i}^{\sigma \dagger} 
         a_{\nu j}^{\sigma} + {\rm H.C.} \bigl) \nonumber \\
&+& \sum_{\sigma} \sum_{\mu} 
         I_{\mu 0}  
         \xi_{\mu 0}^{\sigma \dagger} 
         \xi_{\mu 0}^{\sigma} \nonumber \\
&+& \sum_{\sigma} \sum_{j \ne 0} \sum_{\nu}  
  \bigr( T_{\mu 0 \nu j}  
         \xi_{\mu 0}^{\sigma \dagger} 
         a_{\nu j}^{\sigma} + {\rm H.C.} \bigl)
\label{H_imp} 
\end{eqnarray}
where $a_{\mu i}^{\sigma \dagger}$ ($a_{\mu i}^{\sigma}$) 
is a creation (annihilation) operator of an electron with 
spin $\sigma$ at the $\mu$th Wannier orbitals 
 at the $i$th iron site.  
The quantities $\epsilon_{\mu i}$ and $t_{\mu i \nu j}$ are ionization potential and transfers between iron sites, respectively, which are expressed as 
\begin{eqnarray}
 \epsilon_{\mu i}
 = \langle \phi_{\mu i} | {\cal H} | \phi_{\mu i} \rangle
\ \ {\rm and} \ \   
 t_{\mu i \nu j}
 = \langle \phi_{\mu i} | {\cal H} | \phi_{\nu j} \rangle 
\label{t_pure}
\end{eqnarray}
with $| \phi_{\mu i} \rangle$=$a_{\mu i}^{\dagger} | 0 \rangle$. 
For the third and fourth terms in Eq.~(\ref{H_imp}), $\xi_{\mu 0}^{\sigma \dagger}$ ($\xi_{\mu 0}^{\sigma}$) 
is a creation (annihilation) operator of an electron with 
spin $\sigma$ at the $\mu$th orbital of an impurity site at the origin.  
The parameters $I_{\mu 0}$ and $T_{\mu 0 \nu j}$ describe an impurity-ionization potential and transfers between the impurity and iron sites, which are given by 
\begin{eqnarray}
 I_{\mu 0}
 = \langle \Phi_{\mu 0} | {\cal H} | \Phi_{\mu 0} \rangle
\ \ {\rm and} \ \   
 T_{\mu 0 \nu j}
 = \langle \Phi_{\mu 0} | {\cal H} | \phi_{\nu j} \rangle 
\label{t_imp}
\end{eqnarray}
with $| \Phi_{\mu 0} \rangle$=$\xi_{\mu 0}^{\dagger} | 0 \rangle$. We note that our impurity model in Eq.~(\ref{H_imp}) does not contain the interaction terms. In the present study, we put emphasis on examining impurity effects of non-magnetic impurity. 

A crystal Hamiltonian ${\cal H}_0$ without impurities is given by 
\begin{eqnarray}
{\cal H}_{0}
&=& \sum_{\sigma} \sum_{i} \sum_{\mu} 
         \epsilon_{\mu i}  
         a_{\mu i}^{\sigma \dagger} 
         a_{\mu i}^{\sigma} \nonumber \\ 
&+& \sum_{\sigma} \sum_{i < j} \sum_{\mu\nu}  
  \bigr( t_{\mu i \nu j}  
         a_{\mu i}^{\sigma \dagger} 
         a_{\nu j}^{\sigma} + {\rm H.C.} \bigl).
\label{H_0} 
\end{eqnarray}
In the above, we assume that 
the iron ionization potentials $\epsilon_{\mu i}$ and the transfers between iron sites $t_{\mu i \nu j}$ are the same as those of ${\cal H}$ in Eq.~(\ref{H_imp}).
We define effective onsite impurity potentials measured from the 
Fe-3$d$ levels as follows:
\begin{eqnarray}
 \Delta I_{\mu}
 = I_{\mu 0}  
 - \epsilon_{\mu 0}. 
\label{imp_rlx}
\end{eqnarray}
In practical calculations, since we employ a finite system, the resulting potential should be corrected in terms of a difference in the Fermi levels of the two systems: 
\begin{eqnarray}
 \Delta I_{\mu}^{c} = 
 \Delta I_{\mu} - 
 \Delta E_{{\rm F}}
\label{imp_rlx_corr}
\end{eqnarray}
with 
$\Delta E_{{\rm F}}
=E_{{\rm F}}-E_{{\rm F}}^{0}$, 
where $E_{{\rm F}}$ and $E_{{\rm F}}^0$ are the Fermi levels for the impurity and pure systems, respectively. In addition, we define modulations in transfer amplitudes via Eqs.~(\ref{t_pure}) and (\ref{t_imp}) as 
\begin{eqnarray}
 \Delta t_{\mu 0 \nu j}
 = \bigl| T_{\mu 0 \nu j} \bigr| 
 - \bigl| t_{\mu 0 \nu j} \bigr| 
\label{t_change}
\end{eqnarray}
to see the impurity effect on the offsite parameters.~\cite{Ohe} Notice that the above modulation is defined via absolute values of the transfers. 

The onsite impurity potential are often calculated as~\cite{Kemper,Wang}
\begin{eqnarray}
\Delta \tilde{I}_{\mu} = \langle \phi_{\mu 0} | 
\bigl( {\cal H} - {\cal H}_{0} \bigr)  
| \phi_{\mu 0} \rangle, 
\label{imp_norlx}
\end{eqnarray}
where $| \phi_{\mu 0} \rangle$ are 
the iron Wannier orbitals for the pure system (or atomic orbitals representing basis functions of ${\cal H}_0$). The difference between the above $\Delta \tilde{I}_{\mu}$ and $\Delta I_{\mu}$ in Eq.~(\ref{imp_rlx}) is that the former does not consider relaxation effects of the impurity orbital.
In usual tight-binding formalism, the impurity potential is addressed as a local operator. However, when we employ {\it an ab initio} pseudopotential formalism, we need non-local operators to describe electronic structures. Thus, $\Delta \tilde{I}_{\mu}$ can be decomposed into the local and non-local contributions as follows: 
\begin{eqnarray}
   \Delta \tilde{I}_{\mu}^{c} 
 = \langle \phi_{\mu 0} | \Delta V_{{\rm loc}} | \phi_{\mu 0} \rangle 
 + \langle \phi_{\mu 0} | \Delta V_{{\rm NL }} | \phi_{\mu 0} \rangle 
 - \Delta E_{{\rm F}} 
\label{imp_norlx_corr}
\end{eqnarray}
with 
$\Delta V_{{\rm loc}} = V_{{\rm loc}} - V_{{\rm loc}}^{0}$ 
and 
$\Delta V_{{\rm NL}} = V_{{\rm NL}} - V_{{\rm NL}}^{0}$,  
with the Fermi-level correction $\Delta E_{{\rm F}}$.
We give computational details of the calculation in Appendix A.

\section{Results}

Our {\it ab initio} density-functional calculations were performed with {\it Tokyo Ab initio Program Package}.~\cite{Ref_TAPP} With this program, electronic-structure calculations with the generalized-gradient-approximation (GGA) exchange-correlation functional~\cite{Ref_PBE96} were performed using a plane-wave basis set and the Troullier-Martins norm-conserving pseudopotentials~\cite{Ref_PP1} in the Kleinman-Bylander representation.~\cite{Ref_PP2} Pseudopotentials of transition-metal atoms were supplemented by the partial core correction,~\cite{PCC} which is crucial in describing the low-energy band structures correctly. A 3$\times$3$\times$1 supercell containing 18 transition-metal atoms including one impurity atom was employed. The experimental crystal structure was taken from Ref.~\onlinecite{LaFeAsO} for LaFeAsO and the same geometry was used for all the impurity systems. The cutoff energies in wavefunctions and charge densities were set to 64 Ry and 256 Ry, respectively, and a 3$\times$3$\times$3 $k$-point sampling was employed.  

\begin{figure*}[htb]
	\begin{center}
	\includegraphics[width=0.8\textwidth]{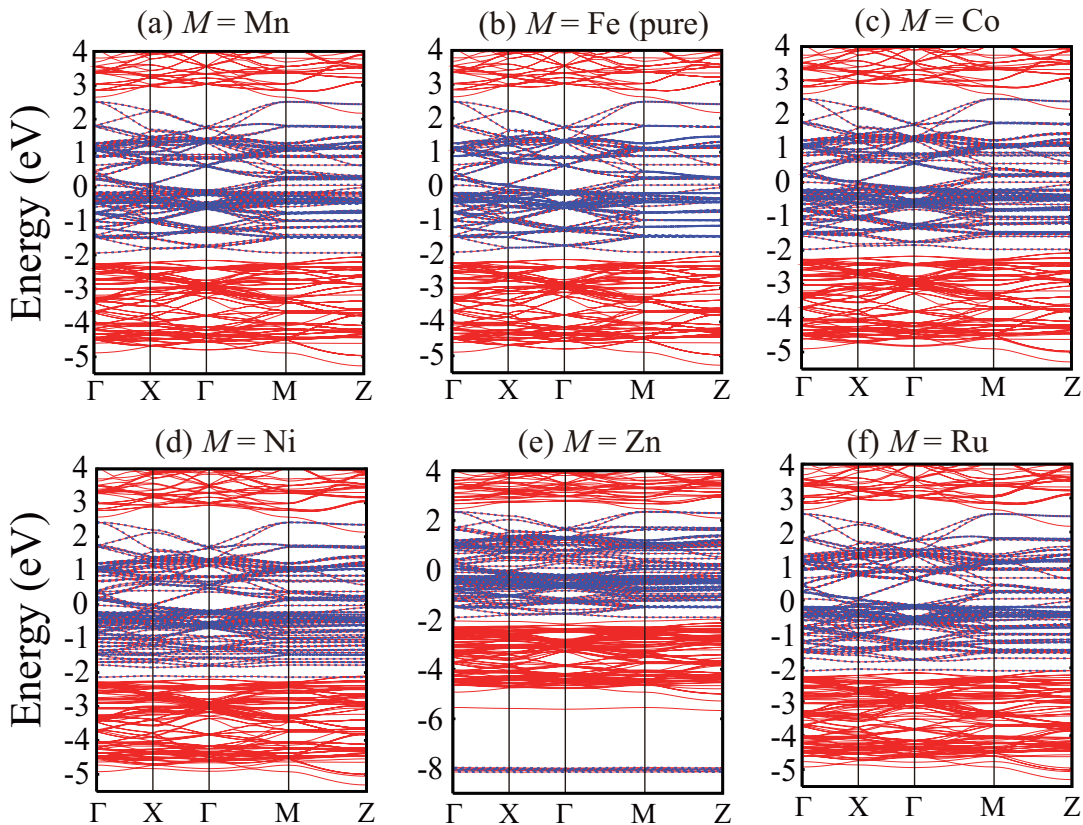}  
	\end{center}
\caption{(Color online)  
Global GGA band structures of ${\rm La} ({\rm Fe}_{0.944}M_{0.056}) {\rm As} {\rm O}$ for a 3$\times$3$\times$1 supercell. (a) $M=$ Mn; (b) Fe (pure); (c) Co; (d) Ni; (e) Zn; (f) Ru. Red-solid and blue-dotted lines are original and Wannier-interpolated band structures, respectively. Band dispersions are plotted along the high-symmetry points, where $\Gamma$ = (0, 0, 0), X = ($a^{*}$/2, 0, 0),
 M = ($a^{*}$/2, $b^{*}$/2, 0), and Z = (0, 0, $c^{*}$/2) and $a^{*}$, $b^{*}$, $c^{*}$ are primitive reciprocal lattice vectors for the supercell.}
\label{Fig1}
\end{figure*}

Figure~\ref{Fig1} shows our calculated supercell GGA band structures of impurity and pure systems, denoted by red solid lines. Tight-binding bands obtained by diagonalization of the impurity Hamiltonian in Eq.~(\ref{H_imp}) are also shown by blue dotted lines, where ionization potentials and transfer integrals [Eqs.~(\ref{t_pure}) and (\ref{t_imp})] were derived with the resulting Wannier functions. The Wannier functions were successfully constructed for Mn, Co, Ni, and Ru, where the $d$ bands are well separated from As-4$p$/O-2$p$ bands, but we failed to do that for Cu even with the ``entangled-band" treatment,~\cite{Souza} because the Cu-3$d$ bands lie deeply inside the As-4$p$/O-2$p$ bands. In the Zn-impurity case [Fig.~\ref{Fig1}~(e)], the Zn-3$d$ levels are much lower than those of the As/O bands, so that it has no trouble to construct the Wannier orbitals. 

To see basic properties of the resulting Wannier functions, we consider the following quantities: 
First, the spatial spread of the Wannier functions in the real space is defined by 
\begin{eqnarray}
  \Omega_{\mu} 
= \sqrt{\langle \Phi_{\mu 0} | r^2 | \Phi_{\mu 0} \rangle
- \bigl| \langle \Phi_{\mu 0} | {\bf r} | \Phi_{\mu 0} \rangle \bigr|^2}.
\label{Spread}
\end{eqnarray}
An averaged value is given as 
$\bar{\Omega} = \bigl(\sum_{\mu}\Omega_{\mu}\bigr)/5$. 
The orbital occupancy of the impurity atom is defined by  
\begin{eqnarray}
 n_{\mu}
 = \sum_{\alpha {\bf k}} f_{\alpha {\bf k}} 
 \bigl| \langle \Phi_{\mu 0} | \psi_{\alpha {\bf k}} \rangle
 \bigr|^2
\label{n_imp}
\end{eqnarray}
with $\psi_{\alpha {\bf k}}$ and $f_{\alpha {\bf k}}$ 
being the Bloch orbital and its occupancy, respectively.  
The site occupancy is given as the sum over the orbitals; 
$n_{{\rm tot}} = \sum_{\mu} n_{\mu}$. 
Overlap integrals between the impurity and pure-Fe orbitals is 
\begin{eqnarray}
 S_{\mu} = \bigl| \langle \Phi_{\mu 0} | \phi_{\mu 0} \rangle \bigr|, 
\label{Overlap}
\end{eqnarray}
which is a quantitative measure of differences in the size and shape of the impurity-$d$ and Fe-$d$ Wannier functions. 
To see how strongly the impurity-$d$ orbitals hybridize with surrounding As orbitals, we consider the center of the Wannier orbitals. From the orbital symmetry, only $d_{yz/zx}$ orbitals exhibit
a finite shift from the iron layer and this shift is calculated as 
\begin{eqnarray}
  \Delta z
= \langle \Phi_{yz/zx 0} | z | \Phi_{yz/zx 0} \rangle - z_{0},
\end{eqnarray}
where $z_0$ denotes the height of the iron layer. 
When $\Delta z$ is negative, the impurity $d_{yz/zx}$ orbitals are close to the As site, indicating a large hybridization.

We summarize in TABLE~\ref{data_Wannier} the results for our calculated Wannier functions. From $\Omega$ and $S$, we see that the size and shape of the Wannier functions of the Mn and Co sites are closely similar to those of the Fe site. In the Ni and Ru cases, $\Omega$ ($S$) is appreciably larger (smaller) than those of Fe, due to stronger hybridization between the impurity-$d$ and the As-$p$ orbitals. Larger $| \Delta z |$ for Ni and Ru than those for Mn, Fe, and Co also supports this observation. On the other hand, for Zn, the $d$ orbitals are almost decoupled with the As-4$p$ orbitals, so that the Wannier functions are as small as those of isolated atomic $d$ orbitals. By comparing calculated site occupancies $n_{{\rm tot}}$ with nominal charges $m$, we see that excess carriers of the impurity are tightly trapped around the impurity site as was shown by Ref.~\onlinecite{Wadati}.
We will discuss these points in more detail in the next section.  
\begin{table*}[htb] 
\caption{A basic property of our calculated maximally localized Wannier orbital (MLWO): Spatial spread $\Omega$ [Eq.~(\ref{Spread})] and orbital occupancy $n$ of MLWO [Eq.~(\ref{n_imp})], and overlap between the impurity orbitals and the iron ones, $S$, defined in Eq.~(\ref{Overlap}) are compared for each element.  An averaged value of the Wannier spread $\bar{\Omega}$, a site occupancy $n_{{\rm tot}}$, and a nominal charge $m$ are also given.  
The bottom row describes a change in the center of the $d_{yz}$ (or $d_{zx}$) MLWO from the iron layer, $\Delta z$. The unit of $\Omega$, $\bar{\Omega}$, and $\Delta z$ are $\AA$. The overlap $S$ is given by \%.}

\
 
\centering 
\begin{tabular}{
c@{\ \ }
r@{\ \ \ }r@{\ \ \ }r@{\ \ }
r@{\ \ \ }
r@{\ \ \ }r@{\ \ \ }
r@{\ \ \ }
r@{\ \ \ }r@{\ \ \ }r@{\ \ }
r@{\ \ \ }
r@{\ \ \ }r@{\ \ \ }r@{\ \ }
r@{\ \ \ }
r@{\ \ \ }r@{\ \ \ }r@{\ \ }
r@{\ \ \ }
r@{\ \ \ }r@{\ \ \ }r} 
 \hline \hline \\  [-2pt]  
       & \multicolumn{3}{c}{Mn} &        
       & \multicolumn{2}{c}{Fe\ \ \ } &
       & \multicolumn{3}{c}{Co} &
       & \multicolumn{3}{c}{Ni} &         
       & \multicolumn{3}{c}{Zn} &       
       & \multicolumn{3}{c}{Ru} \\ [2pt] \hline \\ [-8pt] 
       & $\Omega$\ \ & $n$\ \ & $S$\ \ &   
       & $\Omega$\ \ & $n$\ \ &
       & $\Omega$\ \ & $n$\ \ & $S$\ \ &     
       & $\Omega$\ \ & $n$\ \ & $S$\ \ &    
       & $\Omega$\ \ & $n$\ \ & $S$\ \ &   
       & $\Omega$\ \ & $n$\ \ & $S$
\\ [2pt] \hline \\ [-8pt]
$xy$             & 1.81    & 1.00  & 99.6 &
                 & 1.76    & 1.15  &  
                 & 1.79    & 1.33  & 99.7 &        
                 & 2.66    & 1.54  & 95.9 &      
                 & 0.76    & 2.00  & 82.9 &   
                 & 2.97    & 1.12  & 90.5 \\ 
$yz$             & 2.19    & 0.94  & 99.5 &
                 & 2.12    & 1.17  &  
                 & 2.08    & 1.39  & 99.6 &     
                 & 2.32    & 1.53  & 97.2 &         
                 & 1.11    & 2.00  & 76.0 &
                 & 2.89    & 1.09  & 92.9 \\
$z^2$            & 1.86    & 1.33  & 99.6 &
                 & 1.81    & 1.49  & 
                 & 1.82    & 1.63  & 99.7 &         
                 & 2.20    & 1.72  & 97.1 &   
                 & 0.72    & 2.00  & 83.3 &
                 & 2.62    & 1.43  & 92.7 \\ 
$zx$             & 2.19    & 0.94  & 99.5 &
                 & 2.12    & 1.17  & 
                 & 2.08    & 1.39  & 99.6 & 
                 & 2.32    & 1.53  & 97.2 &  
                 & 1.11    & 2.00  & 76.0 &
                 & 2.89    & 1.09  & 92.9 \\ 
$x^2\!-y^2$      & 2.39    & 0.72  & 99.6 &
                 & 2.37    & 1.01  & 
                 & 2.40    & 1.23  & 99.7 &  
                 & 2.68    & 1.39  & 97.3 &   
                 & 1.18    & 2.00  & 68.0 &
                 & 2.97    & 0.86  & 94.5 \\ [2pt] \hline \\ [-8pt]       
$\bar{\Omega}$   & \multicolumn{3}{c}{2.09} &        
                 & \multicolumn{2}{c}{2.03} &
                 & \multicolumn{3}{c}{2.03} &
                 & \multicolumn{3}{c}{2.44} &         
                 & \multicolumn{3}{c}{0.98} &       
                 & \multicolumn{3}{c}{2.86} \\
$n_{{\rm tot}}$  & \multicolumn{3}{c}{4.93} &        
                 & \multicolumn{2}{c}{6.00} &
                 & \multicolumn{3}{c}{6.96} &
                 & \multicolumn{3}{c}{7.71} &         
                 & \multicolumn{3}{c}{10.0} &       
                 & \multicolumn{3}{c}{5.59} \\ 
$m$              & \multicolumn{3}{c}{5} &        
                 & \multicolumn{2}{c}{6} &
                 & \multicolumn{3}{c}{7} &
                 & \multicolumn{3}{c}{8} &         
                 & \multicolumn{3}{c}{10} &       
                 & \multicolumn{3}{c}{6} \\ 
$\Delta z$       & \multicolumn{3}{c}{   0.002} &        
                 & \multicolumn{2}{c}{$-$0.047} &
                 & \multicolumn{3}{c}{$-$0.097} &
                 & \multicolumn{3}{c}{$-$0.244} &         
                 & \multicolumn{3}{c}{$-$0.063} &       
                 & \multicolumn{3}{c}{$-$0.192} \\ 
\hline \hline \\ 
\end{tabular} 
\label{data_Wannier} 
\end{table*}

We show in TABLE~\ref{IP_change} the onsite-potential differences defined by Eq.~(\ref{imp_rlx_corr}) between the impurity-$d$ and Fe-$d$ orbitals. While the potential difference is positive $\sim$0.3 eV for Mn, those for Co and Ni are negative and estimated to be $\sim-$0.3 eV and $\sim-$0.8 eV, respectively. Since the number of valence $d$ electrons is smaller (larger) for Mn (Ni and Co) and ionization potential of 3$d$-transition metals is deeper for heavier elements, the qualitative tendency of the present result is reasonable. For a reference, we compare the results with ``atomic'' onsite-energy differences from the parameters given by Harrison,~\cite{Harrison} from which we see the same trend. We note that the amplitude of the potential difference should be larger for localized ``atomic" orbitals than for hybridized Wannier orbitals with larger spatial spreads. For the Zn case, we see a rather deep potential,~\cite{Zhang_1} being consistent with the band-structure results of Fig.~\ref{Fig1}~(e). The onsite potential difference is quite small for Ru, since Ru and Fe are isovalent. We note that the Fermi-level correction $\Delta E_{F}$ is very small for all cases (see the bottom row), indicating that the size of the present supercell (3$\times$3$\times$1) is large enough so that interactions between impurities are negligible.
\begin{table}[htb] 
\caption{Calculated onsite-potential differences between the impurity-$d$ and iron-3$d$ orbitals of La(Fe,$M$)AsO, $\Delta I^{c}$, defined in Eq.~(\ref{imp_rlx_corr}).  The supercell size is 3$\times$3$\times$1 with replacing an Fe atom by an impurity $M$ atom.  Averaged values are compared with the results deduced from the Harrison's atomic parameters in Ref.~\onlinecite{Harrison}. The last row is correction due to the Fermi-level difference. 
The unit is eV.}
\ 

\centering 
\begin{tabular}{c@{\ \ \ }r@{\ \ \ }r@{\ \ \ }r@{\ \ \ }r@{\ \ \ }r} 
\hline \hline \\ [-2pt] 
           & Mn   &   Co    &   Ni    &   Zn    &      Ru    \\ 
[2pt] \hline \\ [-8pt] 
$xy$       & 0.32 & $-$0.39 & $-$0.97 & $-$7.88 &   $-$0.10  \\ 
$yz$       & 0.27 & $-$0.34 & $-$0.84 & $-$8.10 &      0.03  \\ 
$z^2$      & 0.29 & $-$0.36 & $-$0.93 & $-$7.96 &   $-$0.21  \\ 
$zx$       & 0.27 & $-$0.34 & $-$0.84 & $-$8.10 &      0.03  \\ 
$x^2\!-y^2$& 0.25 & $-$0.33 & $-$0.75 & $-$8.22 &      0.16  \\ 
[2pt] \hline \\ [-8pt]
Average    & 0.28 & $-$0.35 & $-$0.87 & $-$8.05 &   $-$0.02  \\ 
Ref.~\onlinecite{Harrison}
           & 1.27 & $-$1.23 & $-$2.42 &  -\ \ \ &      1.95  \\ 
[2pt] \hline \\ [-8pt]
$\Delta E_{F}$ & $-$0.004 & 0.011 & 0.032 & 0.084 &  0.062 \\ 
\hline \hline \\ 
\end{tabular} 
\label{IP_change} 
\end{table}

We next show in TABLE~\ref{transfer_change} modulations in diagonal transfer integrals defined by Eq.~(\ref{t_change}). Since we compare the absolute values of the transfers, when the value in the table is positive (negative), the transfers increase (decrease) by impurity doping. For Mn and Co, the modulation is nearly 10 \% compared to the transfers of the pure system (denoted by $t_{0}$ for nearest and $t'_{0}$ for next nearest); for example, for the Mn case, nearest $d_{xy}$-$d_{xy}$ transfer gives $\Delta t/t_{0} \sim 24/313$ and, for next-neighbor $d_{zx}$-$d_{zx}$ transfer, $\Delta t'/t'_{0} \sim 2/337$. For Ni, the modulation becomes larger than the Mn/Co cases ($\Delta t/t_{0} \sim 64/313$ for $d_{xy}$, $\Delta t'/t'_{0} \sim 48/337$ for $d_{zx}$).This is concerned with an increase of the spatial spread of the Ni-3$d$ orbitals (see TABLE~\ref{data_Wannier}). The trend is further enhanced in Ru having more spatially extended 4$d$ orbitals than the Fe-3$d$ ones ($\Delta t/t_{0} \sim 70/313$ for $d_{xy}$, $\Delta t'/t'_{0} \sim 120/337$ for $d_{zx}$).  
\begin{table*}[htb] 
\caption{Calculated modulations in diagonal nearest-neighbor transfers, $\Delta t$, and next-nearest-neighbor ones, $\Delta t'$. The Fe columns include transfers themselves denoted by $t_{0}$ and $t'_{0}$. The unit is meV.}
\
 
\centering 
\begin{tabular}{c@{\ \ \ }r@{\ \ \ }r@{\ \ \ }r@{\ \ \ }r@{\ \ \ }r@{\ \ \ }r@{\ \ \ }r@{\ \ \ }r@{\ \ \ }r@{\ \ \ }r@{\ \ \ }r@{\ \ \ }r@{\ \ \ }r@{\ \ \ }r} 
 \hline \hline \\  [-2pt]  
       & \multicolumn{2}{c}{Mn} &      
       & \multicolumn{2}{c}{Fe} &
       & \multicolumn{2}{c}{Co} &
       & \multicolumn{2}{c}{Ni} &       
       & \multicolumn{2}{c}{Ru} \\ [2pt] \hline \\ [-8pt] 
                 & $\Delta t$\ \ \ & $\Delta t'$\ \ \ &    
                 & $t_{0}$   \ \ \ & $t'_{0}$   \ \ \ & 
                 & $\Delta t$\ \ \ & $\Delta t'$\ \ \ & 
                 & $\Delta t$\ \ \ & $\Delta t'$\ \ \ &     
                 & $\Delta t$\ \ \ & $\Delta t'$\ \ \ \\ [2pt] 
\hline \\ [-8pt]
$xy$             & 23.61    & 1.24     &    
                 & 312.68   & 63.15    & 
                 & $-$26.91 & $-$2.30  & 
                 & $-$63.83 & $-$1.45  &         
                 & 69.72    & 19.53    \\ 
$yz$             & $-$5.95  & 8.04     &   
                 & 215.39   & 148.83   & 
                 & 12.42    & $-$6.72  & 
                 & 55.46    & $-$9.18  &      
                 & 73.21    & 41.16    \\ 
$z^2$            & $-$0.71  & 1.78     &  
                 & 74.24    & 1.97     & 
                 & 4.20     & $-$0.91  &
                 & 28.35    & 0.73     &          
                 & 37.48    & 1.01     \\ 
$zx$             & $-$5.95  & $-$2.12  &  
                 & 215.39   & 336.88   & 
                 & 12.42    & 4.37     & 
                 & 55.46    & 48.40    &  
                 & 73.21    & 120.25   \\ 
$x^2\!-y^2$      & $-$22.51 & $-$3.03  &  
                 & 172.51   & 129.17   & 
                 & 27.78    & 3.41     & 
                 & 84.36    & 15.67    &   
                 & 32.40    & 26.91    \\ [2pt] 
\hline \hline \\ 
\end{tabular}
\label{transfer_change} 
\end{table*}

We give in TABLE~\ref{IP_change_norlx} onsite-potential differences without relaxation of the impurity orbitals, based on Eq.~(\ref{imp_norlx_corr}), together with the decomposition into local $\Delta V_{{\rm loc}}$ and non-local $\Delta V_{{\rm NL}}$ contributions. The value of $\Delta V_{{\rm loc}}$ partially cancels with that of $\Delta V_{{\rm NL}}$ and the subtle balance makes the net value. Basically, for Mn to Ni, the trends of $\Delta \tilde{I}^c$ are the same as those of $\Delta I^c$ including orbital-relaxation effects (shown in TABLE~\ref{IP_change}), although the values themselves are quantitatively different. For Zn (almost atomic 3$d$ orbitals) and Ru (extended 4$d$ orbitals), the orbital relaxation affects the results qualitatively; the sign of $\Delta I^c$ and $\Delta \tilde{I}^c$ are different. 
\begin{table*}[htb] 
\caption{Decomposition of onsite-potential differences between the impurity-$d$ and iron-3$d$ orbitals, without the impurity-orbital relaxation, $\Delta \tilde{I}^c$, into local $\Delta V_{{\rm loc}}$ and non-local $\Delta V_{{\rm NL}}$ contributions, based on Eq.~(\ref{imp_norlx_corr}). See also Appendix A. The unit is eV. }
\   
 
\centering  
\begin{tabular}{
 c@{\ \ \ }
 c@{\ \ \ }c@{\ \ \ }c@{\ \ \ }
 c@{\ }
 c@{\ \ \ }c@{\ \ \ }c@{\ \ \ }
 c@{\ }
 c@{\ \ \ }c@{\ \ \ }c@{\ \ \ }
 c@{\ } 
 c@{\ \ \ }c@{\ \ \ }c@{\ \ \ }
 c@{\ }
 c@{\ \ \ }c@{\ \ \ }c} 
\hline \hline \\ [-2pt] 
 & \multicolumn{3}{c}{Mn} &
 & \multicolumn{3}{c}{Co} &
 & \multicolumn{3}{c}{Ni} & 
 & \multicolumn{3}{c}{Zn} &
 & \multicolumn{3}{c}{Ru} \\ [2pt] 
 \hline \\ [-2pt] 
 & $\Delta V_{{\rm loc}}$ & $\Delta V_{{\rm NL}}$ & $\Delta \tilde{I}^c$ &  
 & $\Delta V_{{\rm loc}}$ & $\Delta V_{{\rm NL}}$ & $\Delta \tilde{I}^c$ &  
 & $\Delta V_{{\rm loc}}$ & $\Delta V_{{\rm NL}}$ & $\Delta \tilde{I}^c$ & 
 & $\Delta V_{{\rm loc}}$ & $\Delta V_{{\rm NL}}$ & $\Delta \tilde{I}^c$ & 
 & $\Delta V_{{\rm loc}}$ & $\Delta V_{{\rm NL}}$ & $\Delta \tilde{I}^c$ \\ 
 [2pt] \hline \\ [-5pt] 
 $xy$    & 4.61    & $-2.99$   &   1.63   & 
 \       & 2.44    & $-2.62$   & $-0.19$  & 
 \       & 3.02    & $-3.40$   & $-0.41$  &  
 \       & $-7.45$ &  9.31     &  1.78    & 
 \       & 1.30    &  14.91    & 16.15    \\
 $yz$    & 4.02    & $-$2.61   & 1.42     & 
 \       & 2.14    & $-2.30$   & $-0.17$  &  
 \       & 2.62    & $-3.00$   & $-0.41$  &   
 \       & $-6.66$ & 8.10      & 1.36     & 
 \       & 1.09    & 13.14     & 14.16    \\  
 $z^2$   & 4.49    & $-$2.94   & 1.56     & 
 \       & 2.38    & $-2.54$   & $-0.17$  &  
 \       & 2.90    & $-3.30$   & $-0.42$  &   
 \       & $-7.47$ & 9.08      & 1.53     & 
 \       & 1.22    & 14.36     & 15.51    \\   
 $zx$    & 4.02    & $-$2.61   & 1.42     & 
 \       & 2.14    & $-2.30$   & $-0.17$  &  
 \       & 2.62    & $-3.00$   & $-0.41$  &   
 \       & $-6.66$ & 8.10      & 1.36     & 
 \       & 1.09    & 13.14     & 14.16    \\  
$x^2\!-\!y^2$& 3.41& $-$2.20   & 1.22     & 
 \       & 1.82    & $-1.98$   & $-0.17$  &  
 \       & 2.24    & $-2.59$   & $-0.38$  &   
 \       & $-5.61$ & 6.85      & 1.16     & 
 \       & 0.93    & 11.42     & 12.29    \\  [2pt]  
\hline \hline \\
\end{tabular}
\label{IP_change_norlx} 
\end{table*} 

\section{Discussion} 

When impurities are doped into superconductors, we can consider two kinds of effects. One is the so-called pair-breaking effect, which can be theoretically studied by onsite-potential differences or transfer modulations in the tight-binding Hamiltonian. As was mentioned above, the chemical trend of these parameters can be understood in terms of the atomic $d$ levels of the impurities; when the $d$ level is low, the potential difference and the transfer modulation due to the hybridization with As-4$p$ orbitals are large. The other impurity effect is carrier doping. As was discussed in Ref.~\onlinecite{Wadati}, theoretically, the excess carriers of the dopants are trapped around the impurity (TABLE~\ref{data_Wannier}), apparently suggesting that the impurities does not supply carriers to the system. On the other hand, many experiments such as the Hall-conductivity measurement~\cite{Sato} suggest that the impurity supplies carriers to the system. Below, we discuss this problem in more details.

\begin{figure*}[htb]
	\begin{center}
	\includegraphics[width=0.8\textwidth]{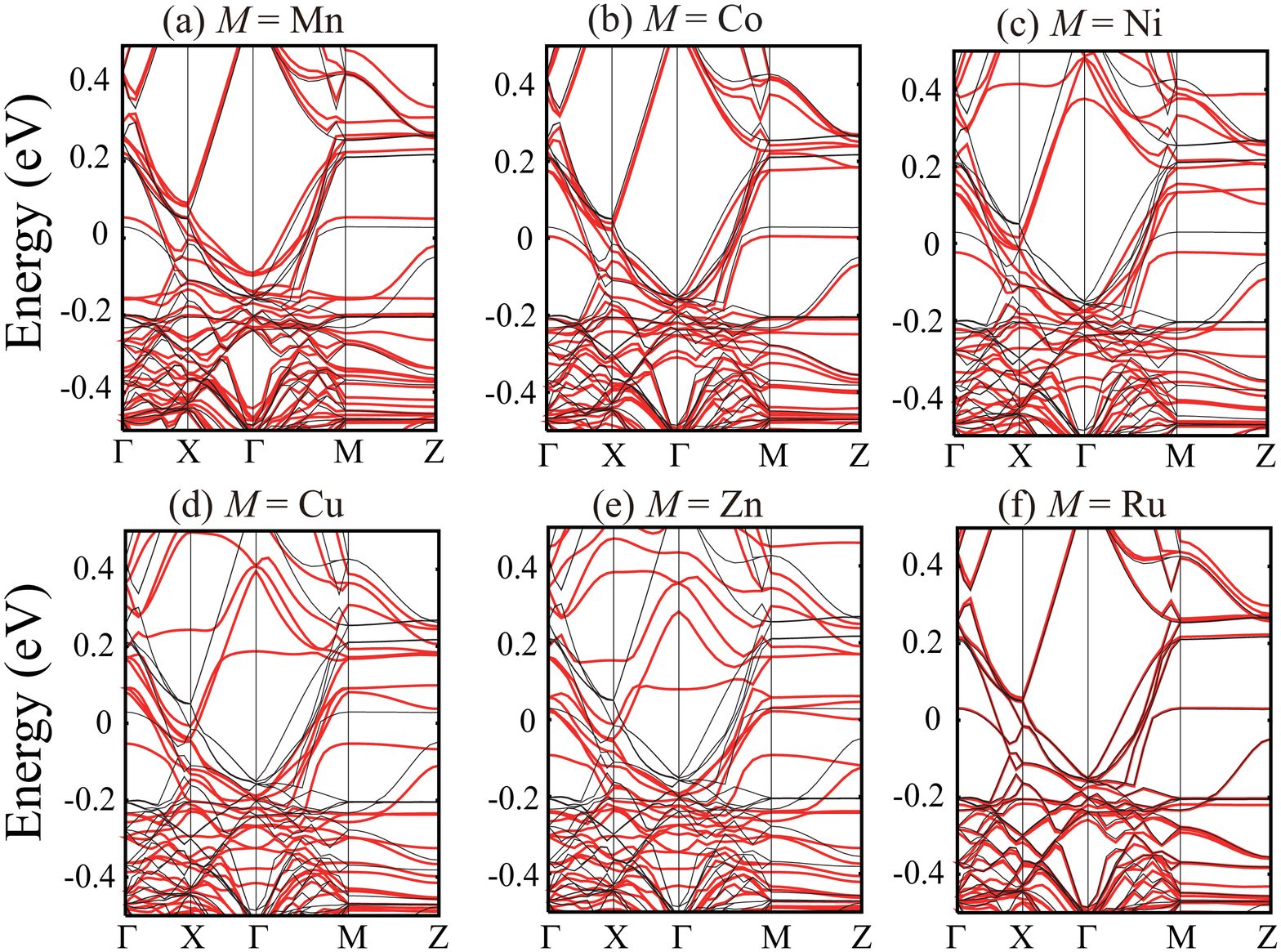}  
	\end{center}
\caption{(Color online) 
Low-energy GGA band structures of ${\rm La} ({\rm Fe}_{0.944}M_{0.056}) {\rm As} {\rm O}$ for the range of [$-$0.5 eV: 0.5 eV]. (a) $M=$ Mn; (b) Co; (c) Ni; (d) Cu; (e) Zn; (f) Ru.
Thick-red and thin solid lines are band structures of the impurity-$M$ and the pure systems, respectively.  }
\label{Fig2}
\end{figure*}

Figure~\ref{Fig2} shows low-energy band structures near the Fermi level. 
To see impurity effects on the band structures, 
we compare the band for the impurity system (thick-red lines) with that for the pure system (thin-solid lines). 
For the cases of Mn, Co, Ni, and Cu, we see a systematic downward rigid shift of the band structure.
For the Zn case, the band structure of the impurity system is totally different from the original band structure, because the Zn-3$d$ orbitals do not exist in the low-energy region [see also Fig.~\ref{Fig1}~(e)]. For the Ru case, we see a similarity between two band structures, implying a weak effect on the low-energy band structures. 

\begin{figure*}[htb]
	\begin{center}
	\includegraphics[width=0.5\textwidth]{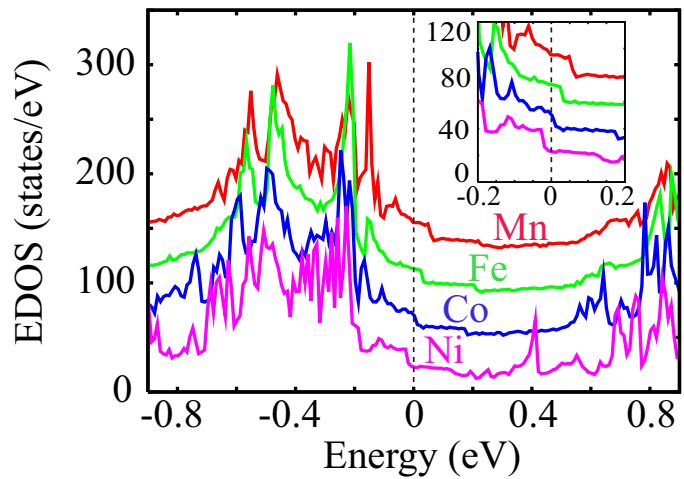}  
	\end{center}
\caption{(Color online) Comparison of electronic density of states (EDOS) for ${\rm La} ({\rm Fe}_{0.944}M_{0.056}) {\rm As} {\rm O}$ with $M$ = Mn (red), Fe (green), Co (blue), and Ni (pink). For each EDOS, offset values of 40 states/eV for Co, 80 states/eV for Fe, and 120 states/eV for Mn are applied. (Inset) Enlarged EDOS for the range of [$-$0.2 eV: 0.2 eV] with offset values of 20 states/eV for Co, 40 states/eV for Fe, and 60 states/eV for Mn.}
\label{Fig3}
\end{figure*}

To show the above observed rigid shift clearer, in Fig.~\ref{Fig3}, we superpose electronic density of states calculated for the Mn- to Ni-impurity cases. The nature of the rigid-band shift well holds especially for the low-energy region near the Fermi level (inset). This rigid shift makes the same changes in the Fermi surface as by doping. For Mn, the electron (hole) pocket is larger (smaller) than that of the pure system. The opposite results are obtained for the Co and Ni cases. We note that the uniform shift in $k$-space band dispersion reflects the locality of the potential change formed around an impurity site.

\begin{figure*}[htb]
	\begin{center}
	\includegraphics[width=0.93\textwidth]{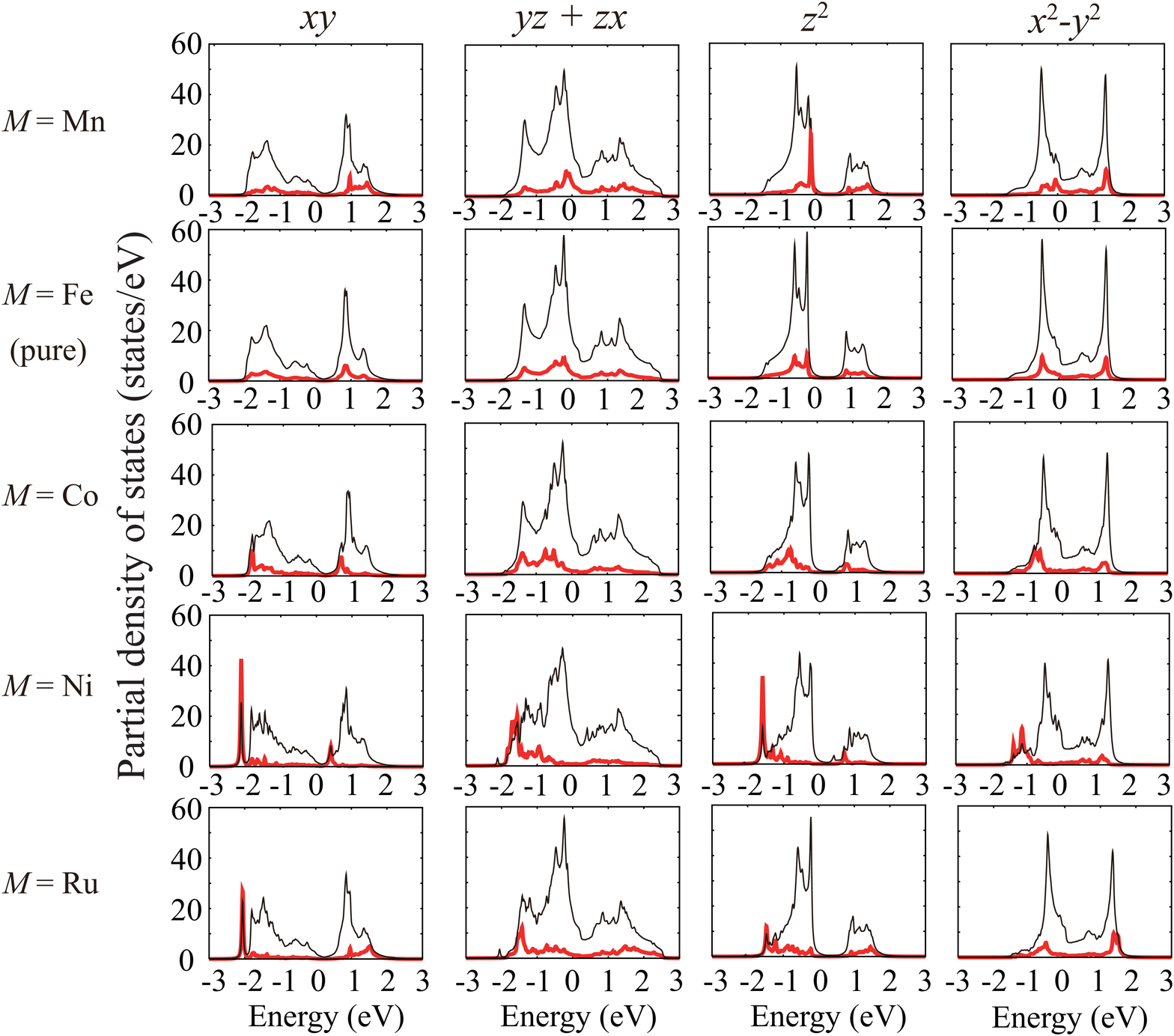}  
	\end{center}
\caption{(Color online) Partial density of states for ${\rm La} ({\rm Fe}_{0.944} {M}_{0.056}) {\rm As} {\rm O}$ of each $d$ orbital. From the top, $M=$ Mn, Fe (pure), Co, Ni, and Ru. Thin-black and thick-red lines are total [Eq.~(\ref{pDOS_total})] and impurity [Eq.~(\ref{pDOS_imp})] spectra, respectively. The impurity spectra are tripled. The Lorentz broadening of 0.02 eV is applied.}
\label{Fig4}
\end{figure*}

In order to examine this point in more details, we calculate partial density of states (pDOS) of the $d$ orbitals. The pDOS is further decomposed into the impurity-$M$ and iron contributions as 
\begin{eqnarray}
 \rho_{\mu}(\epsilon) = \rho_{\mu}^{M}(\epsilon) + \rho_{\mu}^{{\rm Fe}}(\epsilon) 
\label{pDOS_total}
\end{eqnarray}
with 
\begin{eqnarray}
 \rho_{\mu}^{M}(\epsilon)
 = \sum_{\alpha {\bf k}} 
 \bigl| \langle \Phi_{\mu 0} | \psi_{\alpha {\bf k}} \rangle \bigr|^2
 \delta (\epsilon-\epsilon_{\alpha {\bf k}})
\label{pDOS_imp}
\end{eqnarray}
and 
\begin{eqnarray}
 \rho_{\mu}^{{\rm Fe}}(\epsilon)
 = \sum_{i\ne0} \sum_{\alpha {\bf k}} 
 \bigl| \langle \phi_{\mu i} | \psi_{\alpha {\bf k}} \rangle \bigr|^2
 \delta (\epsilon-\epsilon_{\alpha {\bf k}}).
\label{pDOS_Fe}
\end{eqnarray}
Figure~\ref{Fig4} displays the resulting pDOS for Mn (top) to Ru (bottom). 
Thin-black and thick-red solid lines are the total [Eq.~(\ref{pDOS_total})] and impurity [Eq.~(\ref{pDOS_imp})] spectra, respectively. (The latter weight is tripled.)  We see from the figure that, from Mn to Ni, the weight of the impurity pDOS shift from the high-energy to low-energy sides. 

The observation for the rigid shift is quantified by calculating the centers of pDOS in Eqs.~(\ref{pDOS_total})-(\ref{pDOS_Fe}), defined as  
\begin{eqnarray}
& &C_{\mu}^{{\rm tot}}
 = \frac{\int d\epsilon \epsilon \rho_{\mu}(\epsilon)}
        {\int d\epsilon \rho_{\mu}(\epsilon)}, 
\label{center_of_pDOS_tot} \\ 
& &C_{\mu}^{M} 
 = \frac{\int d\epsilon \epsilon \rho_{\mu}^{M}(\epsilon)}
        {\int d\epsilon \rho^{M}_{\mu}(\epsilon)}, 
\label{center_of_pDOS_imp} \\ 
& &C_{\mu}^{{\rm Fe }} 
 = \frac{\int d\epsilon \epsilon \rho_{\mu}^{{\rm Fe}}(\epsilon)}
        {\int d\epsilon \rho_{\mu}^{{\rm Fe}}(\epsilon)}.
\label{center_of_pDOS_Fe}
\end{eqnarray}
TABLE~\ref{pDOS_center} shows calculated centers of pDOS and its decomposition into impurity and iron contributions.
Notice that, for the pure case, $C_{\mu}^{{\rm tot}}$=$C_{\mu}^{M}$=$C_{\mu}^{{\rm Fe}}$, so we give only $C_{\mu}^{{\rm tot}}$ in the table. Clearly, from $M=$ Mn to Ni, the quantity $C_{\mu}^{M}$ shows a systematic downward shift, while the center of the iron pDOS, $C_{\mu}^{{\rm Fe}}$, is nearly unchanged for all the materials. Thus, the shift of $C_{\mu}^{M}$ results in the systematic shift of $C_{\mu}^{{\rm tot}}$.
\begin{table*}[htb] 
\caption{Calculated centers of partial density of states, $C_{\mu}^{{\rm tot}}$ in Eq.~(\ref{center_of_pDOS_tot}), and the decomposition into contributions from impurity [$C_{\mu}^{M}$ in Eq.~(\ref{center_of_pDOS_imp})] and iron [$C_{\mu}^{{\rm Fe}}$ in Eq.~(\ref{center_of_pDOS_Fe})] atoms.}

\
 
\centering 
\begin{tabular}{
c@{\ \ }
r@{\ \ \ }r@{\ \ \ }r@{\ \ }
r@{\ \ \ \ \ }
r@{\ \ \ }
r@{\ \ \ \ \ }
r@{\ \ \ }r@{\ \ \ }r@{\ \ }
r@{\ \ \ \ \ }
r@{\ \ \ }r@{\ \ \ }r@{\ \ }
r@{\ \ \ \ \ }
r@{\ \ \ }r@{\ \ \ }r} 
 \hline \hline \\  [-2pt]  
       & \multicolumn{3}{c}{Mn} &        
       & Fe\ \ \  &
       & \multicolumn{3}{c}{Co} &
       & \multicolumn{3}{c}{Ni} &         
       & \multicolumn{3}{c}{Ru} \\ [2pt] \hline \\ [-8pt] 
       & $C^{M}$\ & $C^{{\rm Fe}}$\ & $C^{{\rm tot}}$\ &   
       & $C^{{\rm tot}}$&       
       & $C^{M}$\ & $C^{{\rm Fe}}$\ & $C^{{\rm tot}}$\ &   
       & $C^{M}$\ & $C^{{\rm Fe}}$\ & $C^{{\rm tot}}$\ &   
       & $C^{M}$\ & $C^{{\rm Fe}}$\ & $C^{{\rm tot}}$\ \\ 
[2pt] \hline \\ [-8pt]
$xy$             & 0.05    &$-$0.27&$-$0.25&
                 &$-$0.27  &   
                 &$-$0.66  &$-$0.28&$-$0.30&        
                 &$-$1.24  &$-$0.30&$-$0.35&      
                 &$-$0.37  &$-$0.27&$-$0.28\\ 
$yz$             & 0.33    & 0.07  & 0.08  &
                 & 0.07    &   
                 &$-$0.27  & 0.06  & 0.04  &     
                 &$-$0.77  & 0.03  &$-$0.01&         
                 & 0.09    & 0.06  & 0.06 \\
$z^2$            & 0.15    &$-$0.14&$-$0.12&
                 &$-$0.14  & 
                 &$-$0.51  &$-$0.15&$-$0.17&         
                 &$-$1.08  &$-$0.18&$-$0.23&   
                 &$-$0.36  &$-$0.15&$-$0.16\\ 
$zx$             & 0.33    & 0.07  & 0.08  &
                 & 0.07    & 
                 &$-$0.27  & 0.06  & 0.04  & 
                 &$-$0.77  & 0.03  &$-$0.01&  
                 & 0.09    & 0.06  & 0.06  \\ 
$x^2\!-y^2$      & 0.50    & 0.25  & 0.26  &
                 & 0.25    & 
                 &$-$0.08  & 0.24  & 0.22  &  
                 &$-$0.50  & 0.22  & 0.18  &   
                 & 0.41    & 0.24  & 0.25  \\ 
[2pt] \hline \\ [-8pt] 
Average          & 0.27    & 0.00  & 0.01  &
                 &$-$0.01  & 
                 &$-$0.36  &$-$0.02&$-$0.03&  
                 &$-$0.87  &$-$0.04&$-$0.09&   
                 &$-$0.03  &$-$0.01&$-$0.01\\  
\hline \hline \\ 
\end{tabular} 
\label{pDOS_center} 
\end{table*}

In summary, according to our presented observations, the impurities are classified as three groups: (i) For Mn, Co, and Ni doping, the rigid-band shift is appreciable and consistent with the fact that these dopants work as effective doping, which has been already verified experimentally.~\cite{Sato} (ii) In contrast to those, the Zn doping generate a large modulation in the low-energy band structure, which will lead to strong pair breaking for superconductivity. Supposing that Cu is in between Ni and Zn, the Cu substation is expected to have the both effects in the same weight. (iii) For analyses of the Ru doping, we need careful considerations; in this system, the transfer modulations are appreciable and a care is required when we solve the impurity model with this transfer modulation. How it works as pair breaking for superconductivity is an interesting future issue. 

Finally, we give a remark for the trend in electron correlations. The {\it ab initio} estimations for the interaction parameters of LaFeAsO were done with several techniques.~\cite{interaction} Basically, the magnitude of the interaction parameters is dominated by two factors; one is the spatial spread of the Wannier orbital and the other is the number of the screening channels. Since the latter is common for all the materials, the Wannier-spread difference makes an effective difference in the interaction values. As far as we see the Wannier spread $\Omega$ listed in TABLE~\ref{data_Wannier}, we expect that (i) the interactions for Mn and Co will be close to those for the pure Fe system because these elements have similar $\Omega$ values, while (ii), as a result of larger spatial spreads of Ni and Ru than the Fe ones, these interactions are smaller than the Fe ones; less correlated.

\section{Conclusion} 
We have calculated parameters specifying a single-impurity Hamiltonian of iron-based superconductors, such as onsite-energy differences between the impurity- and iron-$d$ orbitals and transfer modulations, with maximally localized Wannier orbitals based on {\it ab initio} density functional calculations. The onsite-energy difference for Mn is positive and those for Co and Ni are negative, which are basically understood in terms of differences in atomic 3$d$ levels of each element. For the Zn case, the impurity $d$ levels are no longer near the Fermi level, so, in this system, the impurity behaves as a vacancy. For the Ru case, while the onsite-level difference is quite small, due to more spatial spread of the Ru orbitals than the Fe ones, the transfer changes are appreciable and estimated to be $\sim$70 meV for the nearest neighbors and $\sim$120 meV for the next neighbors. 

We have found that the calculated site occupancy of the impurity is close to a nominal value, so that the impurities apparently do not work as carrier dopants. Nevertheless, an effective doping can occur as a result of a rigid band shift of low-energy electronic structures toward the lower energy side, by a replacement to heavier impurity elements instead of Fe. The origin is understood by the shift of the center of the impurity pDOS, leading to the depression of the weight of the impurity states near the Fermi level. 

Finally, we emphasize that the present discussions are based on an assumption that the impurities are non-magnetic. When we step forward to studies on the correlation effects in the impurity sites, we have to consider the magnetic properties of impurities more seriously. This point is expected to be important for Mn that can have the largest local moment in the 3$d$ transition metals, remained as an important future problem.

\begin{acknowledgements} 
This work was in part supported from MEXT Japan under the grant numbers 22740215, Funding Program for World-Leading Innovative RD on Science and Technology (FIRST program) on ``Quantum Science on Strong Correlation''. All the computations have been performed on Hitachi SR11000 system at the Supercomputer Center, Institute for Solid State Physics, the University of Tokyo and on the same system of Supercomputing Division, Information Technology Center, the University of Tokyo. We thank Yoshihide Yoshimoto, Shun Konbu, Takahiro Misawa, Hiroshi Shinaoka, and Masatoshi Imada for valuable discussions and comments. We are also grateful to Peter Joseph Hirschfeld, Dai Hirashima, Hiroshi Kontani, Seiichiro Onari, Masatoshi Sato, Yoshiaki Kobayashi, Kazuhiko Kuroki, Yuki Nagai, Hideo Aoki, Toshikaze Kariyado, and Masao Ogata for useful discussions. 
\end{acknowledgements}

\appendix
\section{Calculation of $\Delta \tilde{I}^{c}_{\mu}$ in Eq.~(\ref{imp_norlx_corr})}

In density-functional Kohn-Sham (KS) scheme, ${\cal H\!-\!H_{{\rm 0}}}$ 
in Eq.~(\ref{imp_norlx}) 
is replaced by the KS Hamiltonians; 
${\cal H_{{\rm KS}}\!-\!H^{{\rm 0}}_{{\rm KS}}}$. 
Since the kinetic-energy parts in ${\cal H_{{\rm KS}}}$ and 
${\cal H^{{\rm 0}}_{{\rm KS}}}$ cancel with each other, 
the net difference is made by local contributions 
$\Delta V_{{\rm loc}}$=$V_{{\rm loc}}\!-\!V_{{\rm loc}}^{0}$ 
and non-local ones $\Delta V_{{\rm NL}}$=$V_{{\rm NL}}\!-\!V_{{\rm NL}}^{0}$. 
Thus, matrix elements of $V_{{\rm loc}}$ and $V_{{\rm NL}}$ in the Wannier orbitals are basic ingredients to specify $\Delta \tilde{I}^{c}_{\mu}$ in Eq.~(\ref{imp_norlx_corr}). 

The matrix elements of $V_{{\rm loc}}$ is calculated in the real space as 
\begin{eqnarray}
\langle \phi_{\mu 0} | V_{{\rm loc}} | \phi_{\nu {\bf R}} \rangle 
\!=\!\frac{1}{N} 
  \sum_{{\bf k}}^{N} 
  \Biggl( \int_{V} 
         \tilde{u}_{\mu {\bf k}}^{*} ({\bf r}) 
         V_{{\rm loc}}({\bf r}) 
         \tilde{u}_{\nu {\bf k}}({\bf r}) \Biggr) e^{-i{\bf k R}}  \nonumber 
\end{eqnarray}
with $\tilde{u}_{\mu {\bf k}}({\bf r})$=$\Bigl( \sum_{{\bf R}} \phi_{\mu {\bf R}}({\bf r}) e^{-i {\bf k R}} \Bigr) e^{-i{\bf k r}}$ and $N$ and $V$ being the number of sampling $k$ points and the volume of unitcell, respectively. 
The quantity $\tilde{u}_{\mu {\bf k}}({\bf r})$ is given by a unitary transform of the cell-periodic part of the Kohn-Sham wavefunction, $\tilde{u}_{\mu {\bf k}}({\bf r}) = \sum_{\alpha} U_{\alpha \mu}({\bf k}) u_{\alpha {\bf k}} ({\bf r})$ with $U_{\alpha \mu}({\bf k})$ being the unitary matrix obtained from the Wannier-function construction.

The non-local part of the pseudopotential $V_{{\rm NL}}$ is written as 
\begin{eqnarray}
  V_{{\rm NL}}
= \sum_{\tau} \sum_{lm} \sum_{l'm'}
  | \beta_{lm}^{\tau} \rangle  
  D_{lml'm'}^{\tau} 
  \langle \beta_{l'm'}^{\tau}| \nonumber 
\end{eqnarray}
with $|\beta_{lm}^{\tau} \rangle$ being a projector characterized by a site index $\tau$ and angular momentums $l$ and $m$ and $D_{lml'm'}^{\tau}$ being a coefficient representing a scattering amplitude. 
Matrix elements of $V_{{\rm NL}}$ in the Wannier orbitals are calculated as 
\begin{eqnarray}
\langle \phi_{\mu 0} | V_{{\rm NL}} | \phi_{\nu {\bf R}} \rangle
\!&\!=\!&\!\frac{1}{N} 
  \sum_{{\bf k}}^{N} \exp (- i {\bf k R}) \nonumber \\ 
\!&\!\times&\!\!\Biggl(\!\sum_{\tau} \sum_{lm} \sum_{l'm'}
         \rho_{\mu lm}^{\tau *} ({\bf k}) 
         D_{lm,l'm'}^{\tau} 
         \rho_{\nu l'm'}^{\tau}({\bf k})\!\Biggr),  \nonumber 
\end{eqnarray}
where $\rho_{\mu lm}^{\tau *} ({\bf k})$=$\sum_{{\bf G}} \tilde{c}_{\mu {\bf k+G}} \beta_{lm}^{\tau}({\bf k+G})$. 
The data \{$\tilde{c}_{\mu {\bf k+G}}$\} are obtained with the Fast Fourier transform of \{$\tilde{u}_{\mu {\bf k}}({\bf r})$\}.


\end{document}